\documentclass[%
 reprint,
 amsmath,amssymb,
 aps,
]{revtex4-2}

\usepackage{graphicx}% Include figure files
\usepackage{dcolumn}% Align table columns on decimal point
\usepackage{bm}% bold math
\usepackage{svg}
\begin{document}

\preprint{APS/123-QED}

\title{Probing the formation of dark interlayer excitons via ultrafast photocurrent}% Force line breaks with \\

\author{Denis Yagodkin$^1$}
\author{Elias Ankerhold$^{1}$}%
\author{Abhijeet Kumar$^{1}$}
\author{Johanna Richter$^{1}$}
\author{Kenji Watanabe$^{2}$}
\author{Takashi Taniguchi$^{3}$}
\author{Cornelius Gahl$^{1}$}
\author{Kirill I. Bolotin$^1$}
\email{kirill.bolotin@fu-berlin.de}
\affiliation{$^1$Department of Physics, Freie Universität Berlin, Arnimallee 14, Berlin 14195, Germany}%
\affiliation{$^2$Research Center for Functional Materials, National Institute for Materials Science, 1-1 Namiki, Tsukuba 305-0044, Japan}
\affiliation{$^3$International Center for Materials Nanoarchitectonics, National Institute for Materials Science, 1-1 Namiki, Tsukuba 305-0044, Japan}
%TC:ignore

%\affiliation{
%\affiliation{
%\affiliation{%


\date{\today}% It is always \today, today,

\begin{abstract}
Optically dark excitons determine a wide range of properties of
photoexcited semiconductors yet are hard to access via conventional
spectroscopies. Here, we develop a time-resolved ultrafast photocurrent
technique (trPC) to probe the formation dynamics of optically dark
excitons. The nonlinear nature of the trPC makes it particularly
sensitive to the formation of excitons occurring at the femtosecond
timescale after the excitation. As proof of principle, we extract the
interlayer exciton formation time 0.4~ps at 160 $\mu$J/cm$^2$
fluence in a MoS$_2$/MoSe$_2$ heterostructure
and show that this time decreases with fluence. In addition, our
approach provides access to the dynamics of carriers and their
interlayer transport. Overall, our work establishes trPC as a technique
to study dark excitons in various systems that are hard to probe by
other approaches.
%\begin{description}
%use the optional argument of the \verb+\item+ command to give the category of each %item. 
%\end{description}
\end{abstract}

\maketitle

%\tableofcontents

%\section{\label{sec:level1}First-level heading:\protect\\ The line
%break was forced \lowercase{via} \textbackslash\textbackslash}
%TC:endignore
Coulomb-bound electron-hole pairs (excitons) dominate the optical response of 2D semiconductors from the group of transitional metal dichalcogenides (TMDs)~\cite{1}. While early studies focused on optically allowed bright excitons, optically forbidden "dark" excitons are much less studied. The radiative recombination of these  latter excitons is suppressed as they involve states with non-zero total momentum, non-integer total spin, or spatially separated electron and hole wavefunctions~\cite{1,6}.  Due to the weak interaction with light, these states have a long lifetime. Dark excitons are also the lowest energy excitation in many
TMDs~\cite{9}. Because of that, dark exciton
states likely dominate the long-range transport of
excitons~\cite{10,11}, determine temperature-dependent optical
spectra~\cite{12}, and are responsible for long-lived spin
signals in TMDs~\cite{8,13}. Furthermore, dark excitons are
promising for realizing interacting bosonic many-body states including
the Bose-Einstein condensate and excitonic Mott
insulator~\cite{14,15,16}.

Special approaches are required to investigate the properties of the dark states due to their weak interaction with light. For example, time- and
angle-resolved photoemission spectroscopy (trARPES) or spectroscopies in the terahertz and far-infrared frequency ranges have been used to study dark excitons in TMDs and their heterostructures~(HS)~\cite{17,18,19}. These approaches typically require large (hundreds of
$\mu$m$^2$ area) homogenous samples or are performed at room temperature. Another approach to probe dark excitons, time-resolved photoluminescence (trPL)~\cite{15,23}, has a submicron
spatial resolution but features a lower time resolution and does not work for states with vanishingly small oscillator strengths. As a result, many questions related to dark exciton formation, e.g. its timescale or the influence of phonon scattering and electron screening, remain unresolved. 

Time-resolved photocurrent spectroscopy (trPC) has recently emerged as an approach to study optical processes in (2D)
semiconductors~\cite{24,25,26,27}. In trPC, a current across the sample is recorded vs. the time delay between two
light pulses impinging onto it.  Critically, the technique is inherently sensitive to nonlinear processes. The approach applies to devices down to the nm-scale and is compatible with other probes such as magnetic or electric fields, temperature, or strain. Finally, as a transport-based technique, trPC is inherently sensitive to the phenomena at contacts and interfaces. Here, we use the non-linear response of two-color trPC to probe the formation dynamics of dark excitonic species. To test our approach, we interrogate the formation dynamics of the the most studied dark excitons in TMDs: interlayer excitons in MoS$_2$/MoSe$_2$ heterostructures. 
\begin{figure*}
    \includegraphics[width=0.95\linewidth]{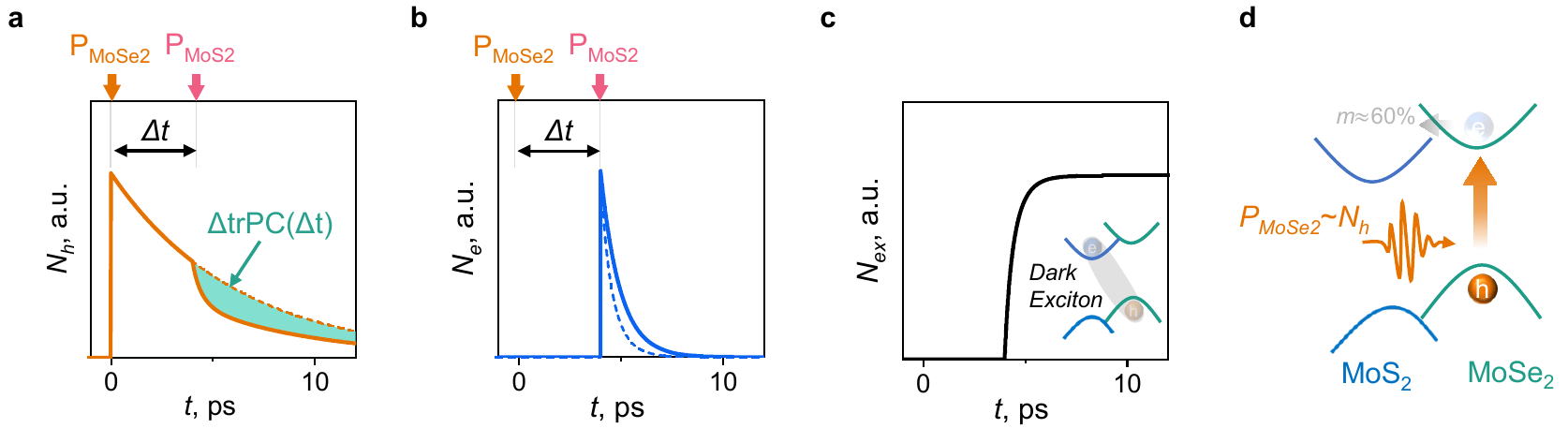}    
    \caption{\textbf{Excitation dynamics and photocurrent}. \textbf{a-c)} Dynamics of holes (a), electrons (b) and excitons (c) modelled by Eq.~1. Solid and dashed lines correspond to non-zero and zero exciton formation rate ($\gamma_{e-h}$). After the excitation by a pulse resonant with the MoSe$_2$ bandgap ($P_{\text{MoSe2}}$) at $t=0$, the hole population ($N_h$) decays exponentially with the rate $\tau_h$. This decay is accelerated after electrons are excited ($P_{\text{MoS2}}$) at $\Delta t\approx \tau_h= 4 $~ps in the case $\gamma_{e-h}\neq0$. The quantitative measure of this acceleration, the shaded area in a, is on the one hand determined by $\gamma_{e-h}$ and on the other hand can be detected in a time-resolved photocurrent (trPC) experiment. \textbf{d)} An optical pump pulse in resonance with MoSe$_2$ bandgap (P$_\text{MoS2}$) excites predominantly holes in the VBM of the heterostructure, while MoS$_2$ resonant pulse excites electrons in CBM (not shown). Binding of electron and hole in individual layers yields dark interlayer excitons (Inset in c).}
    \label{fig:1}
\end{figure*}

{\bf Toy model of time-resolved photocurrent.} Our first goal is to show that the dynamics of dark excitons, which are not accessible to conventional optical techniques, can be obtained from the time-dependent populations of free carriers. To understand this, we consider a simplified model of an optically excited semiconductor. We track the time-dependent densities of free electrons \(N_{e}(t)\) and free holes \(N_{h}(t)\). We assume that electron and hole populations can be excited together (direct excitation) or separately (indirect
excitation), which we model by generation functions \(G_{e}(t)\) and \(G_{h}(t)\). We focus on coupled relaxation
\(\sim N_{e}N_{h}\) which describes the binding of an electron and a hole into an exciton~~\cite{28} (other terms are analyzed in the Supplementary Note 1). Overall, the carrier populations are described within our toy model by the following equations:
\begin{equation}\label{eq:1}
    \begin{cases}
   % \begin{align}
    \dfrac{dN_{e}}{dt} =  - \dfrac{N_{e}}{\tau_{e}} - \gamma_{e - h}N_{e}N_{h} + G_{e} \\
    \\
    \dfrac{dN_{h}}{dt} =  - \dfrac{N_{h}}{\tau_{h}} - \gamma_{e - h}N_{e}N_{h} + G_{h} 
   % \end{align}
    \end{cases}
\end{equation}
Here \(\tau_{e/h}\) are linear decay times of electron and hole
populations, and \(\gamma_{e - h}\) is the nonlinear exciton formation
rate. Since the decay of excitons is several orders of magnitude slower
compared to the decay/trapping of free electrons and
holes~~\cite{29,30}, the density of the excitons is given by 
$$N_{ex}(t) = \int\limits_{- \infty}^{t}{\gamma_{e - h}N_{e}N_{h}dt^{\star}}.$$
While the excitons described by \(N_{ex}(t)\) can be dark (and hence
hard to probe), it can be reconstructed if we have experimental access
to \(N_{e}(t)\) and \(N_{h}(t)\). To accomplish this, we numerically solve the above equations for parameters typical for TMD materials (see SI for details). For
simplicity, we first assume that holes and electrons can be excited
separately. When
only holes are excited at \(t_{1} =0\)~ps and only electrons at, for
example, \(t_{2} \approx \tau_{h} =4\)~ps, the solution yields dynamics
shown in Fig.~1a-c. Initially, the excited population of holes decays
exponentially. After the second pulse arrives, electrons are generated
(Fig.~1b). The decay of holes speeds up due to the formation of excitons
if \(\gamma_{e - h}\) is non-zero. Interestingly, we see that the
population of excitons (Fig.~1c) qualitatively follows the difference
between the hole populations with zero and non-zero
\(\gamma_{e - h}\) (dashed and solid lines in Fig.~1a). For the
non-interacting case (\(\gamma_{e - h} = 0\)), the hole density is not
affected by the second pulse exciting electrons (as in a single pulse
excitation case) and therefore the population of excitons can also be
equated to the difference of hole densities between a single pulse
excitation (\(G_{h} \neq0\) ; \(G_{e} =0\)) vs. two pulse excitation
(\(G_{h}\) \(\neq0\); \(G_{e}\) \(\neq0\)). We see that, in principle,
the dynamics of dark excitons can be obtained from the dynamics of free
carriers.

Two obvious challenges arise when applying this toy model to a realistic
physical system. First, conventional optical techniques, such as
transient reflectivity, detect combined contributions from photoexcited
electron (\(N_{e}\)), hole (\(N_{h}\)), and exciton (\(N_{ex}\))
populations. Second, in conventional semiconductors, optical pulses
generate electrons and holes simultaneously, so the generation functions
\(G_{h}\) and \(G_{e}\) cannot be separately controlled.

To address the first problem, we use time-resolved photocurrent
spectroscopy as our measurement technique. Generally, photocurrent
spectroscopies have the advantage of being directly sensitive to
photogenerated electrons/holes while being insensitive to (neutral)
excitons~~\cite{25,26}. In trPC, the system is illuminated by
two optical pulses separated by the time interval
\(\Delta t\): \(G_{h}\left( t_{0} \right)\) and
\(G_{e}\left( t_{0} + \Delta t \right)\), which generate populations of
\(N_{h}^{0}\) and \(N_{e}^{0}\), respectively. The DC current across the
material is recorded. The trPC signal is defined as the difference in
current with both pulses being present versus only a single pulse. In
general, photocurrent is proportional to the total amount of free
carriers generated in a system over time. For the systems
under study, TMDs, the direct contribution of electrons to the photocurrent can
be neglected as their lifetime is much lower and contact resistance is higher than holes~\cite{25,31} (see Supplementary Note~1). In that case, the photocurrent produced by only the first pulse is
given by the area under the dashed orange curve in Fig.~1a. The decay of
the hole population, in this case, depends only on \(\tau_{h}\) as
\(G_{e} = 0\) for single pulse excitation. The same holds for
\(\gamma_{e - h} = 0\) since the nonlinear term in Eq.~1 vanishes for
both cases. The photocurrent produced by two pulses is given by the area
under the solid curve in Fig.~1a and is smaller than that of a single
pulse because of holes recombining with photoexcited electrons. It can be
shown analytically (Supplementary Note~4) that in the limit of small
\(\gamma_{e - h}\) and \(\tau_{h} \gg \tau_{e}\), the trPC scales
linearly with the exciton formation rate,

\begin{equation}\label{eq:2}
\Delta trPC(\Delta t) \sim \gamma_{e - h}N_{e}^{0}N_{h}^{0}\tau_{e}^{}\tau_{h} \cdot \exp \left (- \frac{\Delta t}{\tau_{{e}/{h}}}\right ) 
\end{equation}

Here the decay time (denominator in the exponent) is \(\tau_{h}\) if holes
are excited first (positive delay time \(\Delta t > 0\)) and
\(\tau_{e}\) if electrons are excited first (negative delay time
\(\Delta t < 0\)). To summarize, trPC can be used to measure both free
carrier parameters (\(\tau_{e},\tau_{h})\) as well as the exciton
formation rate (\(\gamma_{e - h})\).

To address the second problem, we use a 2D heterostructure
MoS$_2$/MoSe$_2$ as a system where the
generation rates for electrons and holes can be controlled separately.
Indeed, the conduction band minimum (CBM) and valence band maximum (VBM)
of the heterostructure reside in different materials,
MoS$_2$ and MoSe$_2$ respectively (Fig.~1d).
Because of that, an optical pulse in resonance with e.g.,
MoSe$_2$ bandgap (${P\textsubscript{MoSe2}}$) excites
holes in the VBM of the structure (MoSe$_2$), while the
excited electrons can relax to the CBM (MoS$_2$) through
tunneling. Crucially, only around \(m \approx 60\%\) of these
electrons reach the CBM of the structure in
MoS$_2$~\cite{21} (Fig.~1d). The remaining electrons are trapped and do not
contribute to photocurrent~\cite{32,33}. These electrons are
not affected by the second pulse~\cite{34}. Similarly, a pulse
resonant with MoS$_2$ bandgap
(${P\textsubscript{MoS2}}$) excites electrons in CBM while
\(m \approx 60 \%\) of holes reach VBM of the structure.
We see that the excitation of MoSe$_2$ produces
predominantly free holes, while the excitation of MoS$_2$
produces predominantly free electrons.

\begin{figure}
    \includegraphics[width=1\linewidth]{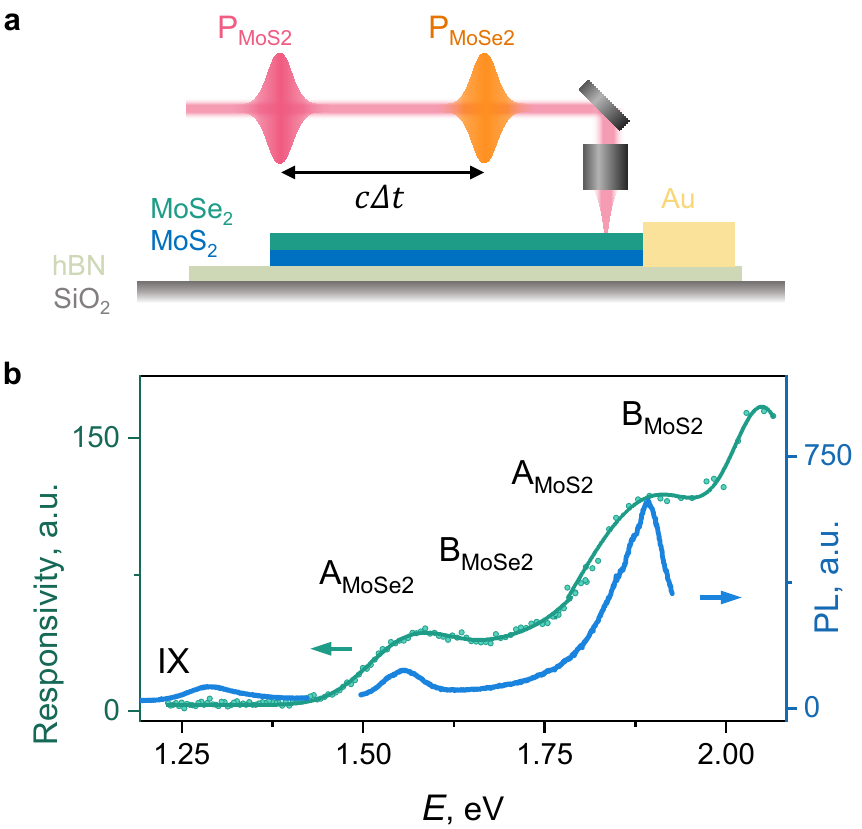}% Here is how to import EPS art
    \caption{\textbf{Sample structure and measurement techniques.} \textbf{a)} Scheme of two-color time-resolved photocurrent measurements (trPC). A photocurrent excited in the TMD heterostructure is measured vs. time delay between a pulse in resonance with MoSe$_2$ ($P_{\text{MoSe2}}$) and a pulse in resonance with MoS$_2$ ($P_{\text{MoS2}}$). \textbf{b)} Photocurrent responsivity (green dots, left axis) and PL (blue line, right axis) spectra  of MoS$_2$/MoSe$_2$ heterostructure. Intralayer A and B excitons of MoS$_2$, MoSe$_2$ are seen (solid green fit) in PC at bias voltage of $3.0 $~V. In PL, an additional feature, an interlayer exciton (IX) is observed.} %Optical image of typical device. \textbf{d)} Static photocurrent map measured at MoSe$_2$ resonance. Photocurrent gives high response in the heterostructure region (grey dashed line) and is greatly amplified near electrodes (yellow dashed rectangles). 
   % \textbf{e)} Photocurrent responsivity spectrum of MoS$_2$/MoSe$_2$ heterostructure (green dots) and MoS$_2$ monolayer (blue dots), bias voltage: $V_b = 3.0 $~V.  Only MoS$_2$-related peaks are present when PC is measured in the monolayer MoS$_2$ region of the same sample (blue dots).}
    \label{fig:2}
\end{figure}

We now apply Eq.~1 to model the excitation dynamics of the
MoS$_2$/MoSe$_2$ heterostructure. The
parameter \(N_{e}(t)\) describes the electron density in the CBM of the
structure (MoS$_2$) and \(N_{h}(t)\) -- the hole density in
the VBM (MoSe$_2$). The free electron/hole decay time (the
term linear with \(N_{e/h}\)) describes the combined contributions of defect capture~\cite{35}, intervalley
scattering~\cite{36}, and radiative decay
processes~\cite{32,33}.
The rate \(\gamma_{e - h}\) describes the formation of (dark) interlayer
excitons. Of course, intralayer excitons are also formed by optical
pulses. However, optically excited intralayer excitons decay much faster 
(within $<100$~fs\cite{17}) compared to intralayer exciton recombination (\textgreater~ps) and electron/hole population cooling rate~\cite{37,38,39} via charge separation across the heterostructure and are therefore subsume in the generation functions $G_e$ and $G_h$. The latter contain contributions from both pulses (i.e.,
\(G_{e}(t) = P_{\text{MoS2}}\left( t,t_{0} \right) + m \cdot P_{\text{MoSe2}}(t,t_{0} + \Delta t)\), see effect of $m$ on dynamics of charge carriers in Fig.~S5). Finally, the excitonic ground state of
MoS$_2$/MoSe$_2$ is an interlayer exciton
comprised of an electron in MoS$_2$ bound to a hole in
MoSe$_2$ (Fig.~1f). When the twist angle between the
heterostructure layers (\(\theta\)) is non-zero, the interlayer
exciton has large in-plane momentum: \(k \sim \frac{\theta}{a}\), where
\(a\) is the averaged lattice constant of the
heterostructure\cite{40,41}. In this case, the radiative
recombination must involve a phonon and the state is
dark\cite{41}. Thus, the interlayer exciton decay
(\textgreater{} 100~ps) can be neglected on the time scales of the
population build-up~\cite{42}. Our next goal is to obtain the dynamics of \(N_{ex}(t)\) via
trPC.

{\bf Time-resolved photocurrent.} For trPC measurements, we fabricate samples, MoS$_2$/MoSe$_2$ on hBN, (Fig.~2a; see Supplementary Note~2 for details). We observe the characteristic intralayer A and B excitons for both materials in static photocurrent spectroscopy in the heterostructure region (green dots in Fig.~2b and Fig.~S1). In addition, weak photoluminescence (PL) due to interlayer excitons (IX) is observed at 1.3~eV~\cite{43} (blue line in Fig.~2b). For time-resolved photocurrent (trPC) measurements, the sample is illuminated with two time-relayed (with $\sim$10~fs precision) optical pulses, one in resonance with the MoS$_2$
bandgap and another with the MoSe$_2$ bandgap. The photocurrent is measured with lock-in amplifier synchronized to an optical chopper in one of the beam paths with no bias voltage applied. This measurement effectively allows us to evaluate the difference between single- and two-pulse responses, which corresponds to the
area between dashed and solid curves in Fig.~1a.

\begin{figure*}
    \includegraphics[width=1\linewidth]{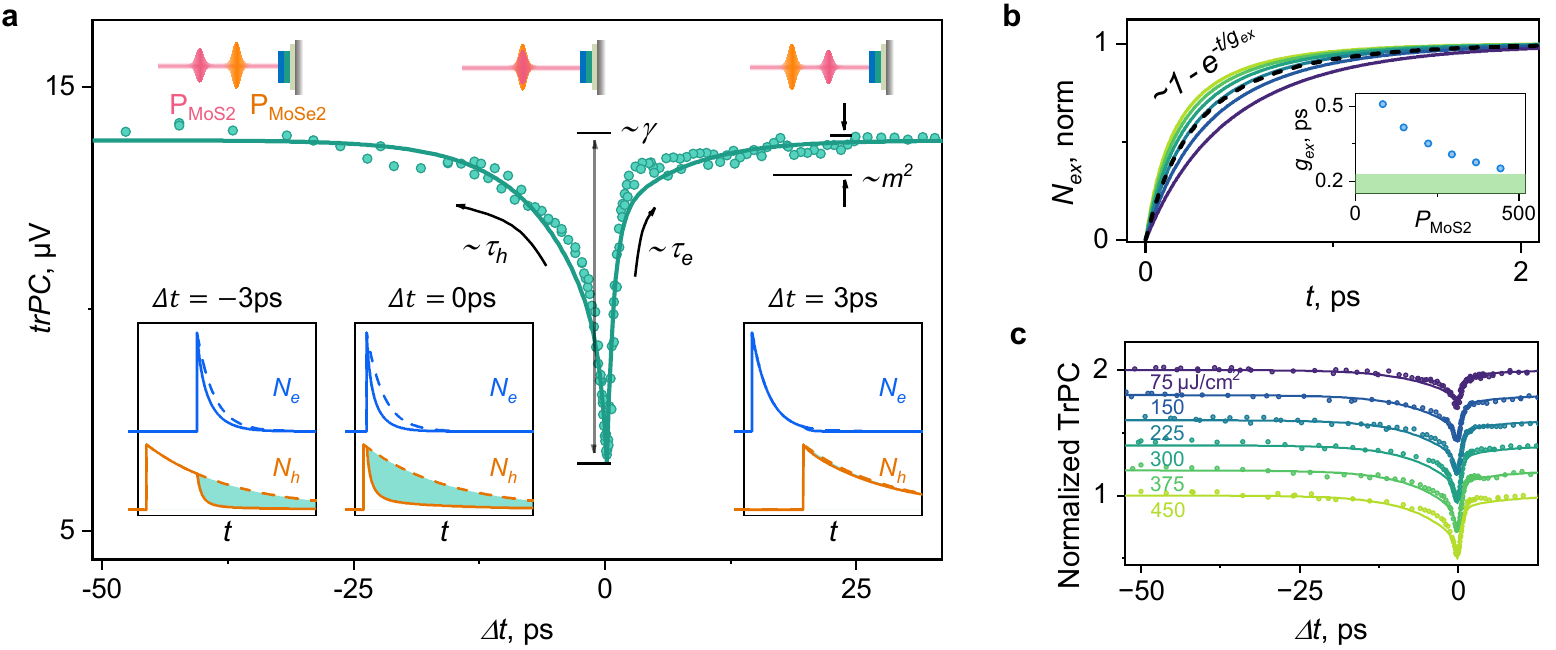}% 
    \caption{\textbf{trPC measurements and extraction of exciton dynamics. a)} trPC response of a MoS$_2$/MoSe$_2$ heterostructure (points). Solid line shows simulated dynamics from Eq.~1 model with following parameters: $\tau_e=1.0 $~ps, $\tau_h=6.0 $~ps, interaction strength $\gamma_{e-h}= 0.13 $~cm$^2$/s, electron/hole tunneling $m=55 \%$. The insets display dynamics of holes (orange) and electrons (blue) for selected delays between pulses for the simplified case $m= 0$. The difference between response to a single optical pulse (solid) and two pulses (dashed) is proportional to $\Delta $trPC (green area). \textbf{b)} Simulated dynamics of interlayer excitons formation for data in a (black dashed line) and for other fluences $P_{\text{MoS2}}  = 75, 150, 225, 300, 375, 450$ $\mu$J/cm$^2$ (solid lines from purple to yellow). For higher fluences exciton formation becomes faster, as quantified by fitting to $1-e^{-t/g_{ex}}$, where $g_{ex}$ is the exciton formation time. Inset:  extracted $g_{ex}$ for the above mentioned fluence range (blue points), green area shows cross correlation of the pulses. The formation time drastically decreases for high laser fluences. \textbf{c)} Normalized fluence dependence of trPC response (points, each dataset is offset by 0.25) and independent from measurement simulations using Eq.~1 with parameters from Fig.~3a (lines). At higher fluences (yellow) the trPC drop at zero-time delay increases, suggesting faster formation and higher number of excitons.}
    \label{fig:3}
\end{figure*}

Figure 3a shows experimental trPC data (green dots) of the
MoS$_2$/MoSe$_2$ sample vs. delay \(\Delta t\)
between the excitation pulses. Positive delay corresponds to the pulses resonant with the MoS$_2$ bandgap arriving first. The most prominent features of the data are a strong dip at zero time delay and a pronounced asymmetry between positive and
negative delays. We now show that these features can be understood
within our toy model. First, a large drop in trPC suggests non-linear
interaction between the carrier populations produced by both pulses,
described by \(\gamma_{e - h}\) in our model (Eq.~2). Second, the
asymmetry can be understood from Eq.~2. It suggests that the lifetime of
electrons photoexcited in MoS$_2$ is much smaller than the
lifetime of holes excited in MoSe$_2$ (see insets in Fig.~3a
for the illustration of trPC at negative, zero, and positive time
delays). Third, we note a slower decaying component at \(\Delta t >3\)~ps.
This minor effect is missing in Fig.~1a and occurs for \(m \neq 0\)
(Supplementary Note~5). Its origin is the transfer of holes to the VBM
(MoSe$_2$). The ratio between fast and slow decaying
components is proportional to \(m^{2}\) (Eq.~S12).

To obtain the precise values of the model parameters, we match the
numerical solution of Eq.~1 (the solid line in Fig.~3a) with the
experimental data. We obtain decay times
\(\tau_{h} = 6.0\pm0.5\)~ps, \(\tau_{e} =1.0 \pm 0.2\)
~ps, interaction strength \(\gamma_{e - h} = 0.13 \pm 0.04\)~cm$^2$/s, and transfer efficiency
\(m = 55\pm5\%\). The effect of every parameter is shown in Fig.~S7. Using the extracted parameters, we plot the generation dynamics of interlayer 
excitons (black dashed line in Fig.~3b). Since the exciton formation is a nonlinear process, its acceleration is expected with a higher density of electrons/holes
which matches simulations at higher fluence (solid lines Fig.~3b). We also extract the dynamics of electrons and holes transferred after photoexcitation within our model (Fig.~S2a). These transferred carriers can be measured with two-color time-resolved reflectivity (trRef) measurements (Supplementary Note 3).
Overall, the simulations suggest two key behaviors. First, we see a much
faster decay rate for electrons compared to holes. Second, we obtain a
formation time of the exciton, $g_{ex} =0.4$~ps, at our experimental incident fluence
\(P_{\text{MoS2}} = 160\)$~\mu$J/cm$^2$. This time is expected to be strongly fluence dependent, dropping by a factor of three for tripled
fluence of ${P\textsubscript{MoS2}}$ (inset in Fig.~3b). 

Next, we test the predictions of these simulations. To independently check the dynamics of free carriers, we carry out two-color time-resolved
reflectivity~\cite{30} (Supplementary Note 3). The resulting time constants of electrons and holes,
\(\tau_{e} =1.4\)~ps and \(\tau_{h} =6.3\)~ps, are close to what is
obtained from trPC. Moreover, our model for trRef suggests that the formation of dark excitons is the
reason behind the bi-exponential decay observed by us and in
other works~\cite{30,31}. To check the dependence of exciton formation on fluence given by our model
(Fig.~3b), we carried out fluence-dependent trPC measurements. We keep
the incident fluence of ${P\textsubscript{MoSe2}}$ fixed, while
${P\textsubscript{MoS2}}$ is varied in the range used in simulation: $75
$~$\mu$J/cm$^2$ -- $450$~$\mu$J/cm$^2$ (dots in Fig.~5).
We see that the magnitude of the drop of the trPC at zero time
delay increases with fluence, from around $30\%$ at $75$
$\mu$J/cm$^2$ up to $50\%$ at $450$~$\mu$J/cm$^2$. The
experimental data closely follow the independent predictions of the
model (lines in Fig.~3c) where we used the same parameters as Fig.~3a
and only changed the fluence of the beams. We extract the formation time of interlayer excitons \(g_{ex}\) (inset in Fig.~3b) and find that it
is more than halved in the given fluence range from 0.5~ps to 0.2~ps.

{\bf Conclusion and outlook.} To summarize the discussion above, the proposed model matches all
observed features of the trPC behavior. We extract the parameters of the
model: e-h coupling
(\(\gamma_{e - h} =0.13 \pm 0.04\)~cm$^2$/s),
decay times of electrons and holes (\(\tau_{e} =1.0 \pm 0.2\)~ps and
\(\tau_{h} =6.0 \pm 0.5\)~ps), and efficiency of interlayer transport
(\(m = 55 \pm 5 \%\)), for all fluence regimes. The interlayer exciton formation time varies from 0.2~ps to 0.5~ps in the range of fluences 450~$\mu$J/cm$^2$ to 75~$\mu$J/cm$^2$. It is useful to compare these values
with those obtained by other approaches. Electron/hole lifetimes are consistent with those broadly reported from optical
measurements~\cite{8,32,39,46,47}. The shorter lifetime of
electrons is likely related to defect states being closer to the
conduction band~\cite{48,49}. The observed exciton formation
time \(g_{ex} = 0.4\)~ps at 160 $\mu$J/cm$^2$ fluence matches
the time scales reported in trARPES ($\sim$230
fs)~\cite{17}, trTHz reflectivity ($\sim$350
fs)~\cite{20}, and trFIR ($\sim$800
fs)~\cite{22} experiments. The interlayer transfer efficiency
\(m\) has been estimated from THz measurements to be
50--70\%~\cite{21}, also close to the values here. Overall,
our approach provides simple access to the dynamics of (dark) interlayer
excitons. Moreover, the proposed model describes trRef dynamics of
heterostructures and explains the biexponential decay reported
before~\cite{30,31}.

To conclude, we demonstrate an approach for studying dark exciton formation dynamics.  In future, this approach can be used to study other dark excitons in TMDs. Unlike other approaches, trPC is fully compatible with other optical techniques (trRef and PL shown here,
Kerr and ellipticity spectroscopies, second harmonic generation), has hundreds of nanometers spatial resolution, and works at cryogenic temperatures. It will be particularly interesting to use trPC to uncover
the effects of many-body interaction (exciton Mott transition, localization at low temperature), electric field, and twist angle on the exciton formation time.
%TC:ignore

\textbf{Acknowledgment}

The authors thank Nele Stetzuhn for her comments on the paper. The authors acknowledge the German Research Foundation (DFG) for financial support through the Collaborative Research Center TRR 227 Ultrafast Spin Dynamics (project B08).

\textbf{Conflict of Interest}

The authors declare no conflict of interest.

\textbf{Author Contribution}

D.Y., K.I.B., and C.G. conceived and designed the experiments, D.Y.,
E.A., A.K., and J.R. prepared the samples, D.Y., A.K. and E.A. performed
the optical measurements, D.Y. analyzed the data, E.A. wrote software
for simulations, D.Y., E.A., performed the calculations and help to
rationalize the experimental data, D.Y. and K.I.B. wrote the manuscript
with input from all co-authors.

\textbf{Data Availability Statement}

The data that support the findings of this study are available from the
corresponding author upon reasonable request.

\bibliography{Exported_Items}
%TC:endignore
\end{document}